\documentclass[pdflatex,sn-mathphys-num]{sn-jnl}

\usepackage{graphicx}%
\graphicspath{{figs/}}%
\usepackage{multirow}%
\usepackage{amsmath,amssymb,amsfonts}%
\usepackage{amsthm}%
\usepackage{mathrsfs}%
\usepackage[title]{appendix}%
\usepackage{xcolor}%
\usepackage{textcomp}%
\usepackage{manyfoot}%
\usepackage{booktabs}%
\usepackage{algorithm}%
\usepackage{algorithmicx}%
\usepackage{algpseudocode}%
\usepackage{listings}%
\usepackage{url}%

\theoremstyle{thmstyleone}%

\theoremstyle{thmstyletwo}%

\theoremstyle{thmstylethree}%

\begin{document}

\title[Automated Patient-Specific EAM Fitting]{Automated Optimization of Patient-Specific Electro-Anatomical Models Fitted to Cochlear Implant SCINSEV}

\author*[1]{\fnm{Minh} \sur{Vu}}
\email{minh.q.vu@vanderbilt.edu}

\author[2]{\fnm{Erin} \sur{Bratu}}

\author[1]{\fnm{Jack} \sur{Noble}}

\affil*[1]{%
\orgdiv{Department of Electrical and Computer Engineering},
\orgname{Vanderbilt University},
\orgaddress{\city{Nashville}, \state{TN}, \country{USA}}}

\affil[2]{%
\orgdiv{INSERM, LTSI -- UMR 1099},
\orgname{University of Rennes},
\orgaddress{\postcode{35000}, \city{Rennes}, \country{France}}}

\abstract{

\textbf{Purpose:}
Patient-specific electro-anatomical models (EAMs) of the implanted cochlea are crucial for auditory nerve fiber simulation in cochlear implant (CI) programming and neural health estimation. Fitting an EAM to the stimulation current induced non-stimulating electrode voltage (SCINSEV) currently requires manual expert adjustment or exhaustive search over precomputed forward solutions, neither of which scales to routine clinical use. We present an automated optimization pipeline that fits patient-specific EAMs in minutes.

\textbf{Methods:}
Four-compartment tissue label maps (perilymph, soft tissue, modiolus, and bone) derived from postoperative CT were used to construct a finite-difference forward model, fitted to the measured SCINSEV by Gaussian-process Bayesian optimization. Three reductions cut fitting cost: a coarsened grid with trilinear injection and readout at sub-voxel contact positions; solving half the stimulating contacts and completing three quarters of the matrix by reciprocity; and a Jacobian-based identifiability analysis reducing four resistivities to two. Each reduction was validated separately against the measurement's reciprocity residual and the reproducibility of the fit.

\textbf{Results:}
Across 43 implanted ears, fitting took a median of 2.6 minutes per ear. Median SCINSEV RMSE on the reduced-resolution objective was 0.069~mV, against 0.084~mV for grid search at a matched evaluation budget. Against an exhaustive full-resolution grid search requiring a median of 9.1 hours per ear, the pipeline reached statistically indistinguishable fitting error roughly $210\times$ faster.

\textbf{Conclusion:}
Patient-specific cochlear EAM fitting can be performed at clinically practical timescales without expert supervision, removing a major computational barrier to patient-specific cochlear implant programming and cochlear simulation.

}


\maketitle

\section{Introduction}

Cochlear implants (CIs) restore hearing by directly stimulating the auditory nerve in individuals with severe to profound hearing loss. Although modern systems provide excellent speech perception for many recipients, outcomes remain highly variable~\cite{ci-outcomes}, motivating continued efforts to personalize device programming. Image-guided cochlear implant programming~\cite{igcip} incorporates patient-specific anatomy into these decisions through computational modeling of intracochlear stimulation.

A key source of patient-specific electrical information is the stimulation current induced non-stimulating electrode voltage (SCINSEV), an objective measurement routinely acquired from contemporary implants~\cite{efi-clinical}. The measurement is named differently by each manufacturer, as the transimpedance matrix (TIM) by Cochlear, electric field imaging (EFI) by Advanced Bionics, and impedance field telemetry (IFT) by MED-EL. We use SCINSEV throughout as the vendor-neutral term. Current is delivered through one stimulating electrode while voltages are recorded at the remaining contacts, producing a matrix that characterizes intracochlear current spread. Because SCINSEV reflects both the electrode configuration and the surrounding conductive anatomy, it provides an objective target for patient-specific modeling. By fitting model parameters so that the simulated SCINSEV matches the measurement, electro-anatomical models (EAMs) can estimate the electrical environment within an individual cochlea without invasive measurement.

Patient-specific EAMs are commonly constructed by combining anatomy from postoperative computed tomography (CT) with SCINSEV~\cite{eam-prior}. Following segmentation of the cochlea and electrode array, tissue compartments are assigned conductivities, intracochlear current propagation is simulated, and the simulated voltages at the contacts are compared with the measured SCINSEV. Parameters are updated iteratively until the two agree. The resulting model supports downstream analyses including current spread characterization, neural activation modeling~\cite{anf-model}, and image-guided programming.

Although conceptually straightforward, this fitting is computationally demanding. Each iteration requires solving the three-dimensional Poisson equation governing the intracochlear potential, typically by the finite-difference method (FDM)~\cite{fdm-cochlea}. Because optimization requires repeated forward simulations, the cost of a single fit is set by the number of forward solves it takes. In this work an exhaustive full-resolution grid search over two parameters required a median of 9.1 hours per ear (Section~\ref{sec:res-overall}). Cost grows further with solver tolerance, model complexity, and the dimension of the search. This burden has limited routine clinical adoption and constrained large-scale retrospective studies.

We present an automated optimization pipeline addressing the major computational bottlenecks in this workflow. It builds on the fitting pipeline described previously~\cite{bratu-lineage}, in which parameters were selected by exhaustive search over a set of precomputed forward solutions. The framework combines a reduced-resolution forward model with sub-voxel electrode interpolation, exploitation of measurement reciprocity, and Bayesian optimization over a reduced parameter set. Evaluated on clinical SCINSEV from 43 implanted ears, it reduced fitting time to a median of 2.6 minutes per ear. Its fitting error was statistically indistinguishable from that of a full-resolution exhaustive grid search requiring a median of 9.1 hours.

\section{Methods}

\begin{figure}[h]
\centering
\includegraphics[width=\textwidth]{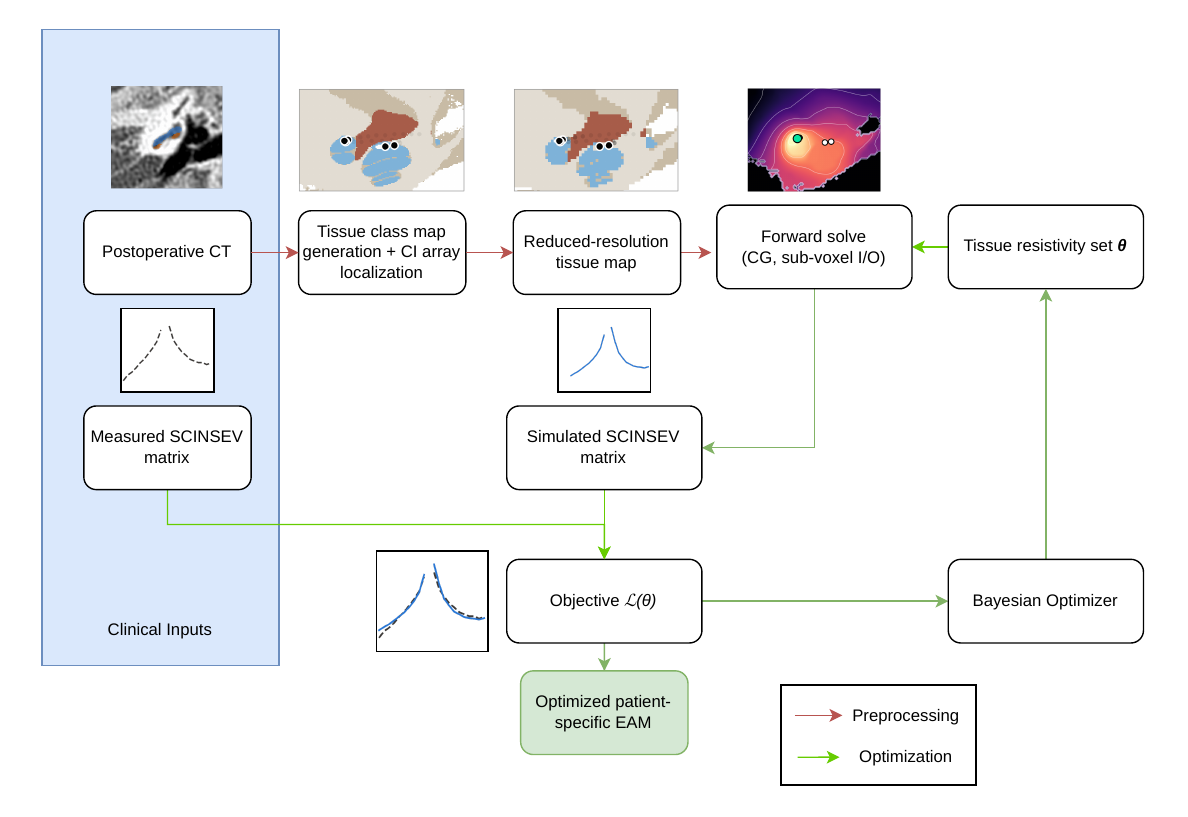}
\caption{Overview of the fitting pipeline. Tissue map construction and array
localization are performed once per ear and amortized across all subsequent
fits; the lower loop is what runs per fit. Each iteration proposes a
resistivity vector, solves the forward model at reduced resolution, and scores
the simulated SCINSEV against the clinical
measurement}\label{fig:pipeline}
\end{figure}

The pipeline takes a postoperative CT and a measured SCINSEV matrix and returns a
fitted patient-specific EAM, with no operator intervention between the two.
As Fig.~\ref{fig:pipeline} shows, tissue map construction and array
localization are performed once per ear. Each fitting iteration then proposes
a resistivity vector, solves the forward model at reduced resolution, and
scores the result against the measurement. The sections below
describe the model formulation, the measurement used as the fitting target,
the optimization procedure, and the experimental evaluation.

\subsection{Patient-specific EAM formulation}

\subsubsection{Tissue map construction}

An EAM is a patient-specific 3D volume conduction model of the implanted cochlea, discretizing the cochlea and surrounding anatomy into a voxel grid in which each voxel carries an electrical resistivity according to its tissue type. Given a current injection at a contact, this permits simulation of the resulting potential distribution.

Tissue labels are assigned by registering high-resolution micro-CT volumes of cadaveric temporal bones to the patient's postoperative CT and determining voxel-wise labels by majority voting across the registered atlas set~\cite{atlas-seg}. The electrode array is localized in the same CT~\cite{elec-localization}, giving the position of each contact within the labeled volume.

The map partitions the domain into four tissue classes: the perilymph-filled scala tympani and scala vestibuli (ST/SV), the modiolus, soft tissue, and bone. Each is treated as homogeneous and isotropic with a single resistivity. This substantially simplifies anatomy that is in reality heterogeneous and anisotropic. Each fitted resistivity is therefore best read as a modeling parameter that reproduces the measured electrical behavior of a region, not as a measurement of tissue conductivity. How the four resistivities are treated during fitting is described in Section~\ref{sec:param-reduction}.

The volume is represented on a regular grid of $251 \times 251 \times 151$ voxels with isotropic spacing of 0.072~mm~\cite{prior-eam-solver}, extending beyond the cochlea to include surrounding bone and soft tissue. Stimulation is applied as a current source at the contact location (Section~\ref{sec:forward-model}). The silicone carrier is represented as an insulating region and excluded from conduction.

\subsubsection{Forward model}
\label{sec:forward-model}

Current flow is modeled as steady-state conduction in a resistive medium. Neglecting capacitive effects~\cite{quasistatic}, the potential $\Phi$ satisfies
\begin{equation}
\nabla \cdot \left( \sigma \nabla \Phi \right) = -I ,
\end{equation}
where $\sigma$ is the spatially varying conductivity given by the inverse of the resistivity map and $I$ the injected current density.

The equation is discretized by the finite-difference method. Adjacent voxels $i$ and $j$ are connected by a conductance
\begin{equation}
g_{ij} = \frac{k}{\rho_i + \rho_j},
\end{equation}
with $k$ a geometric factor set by the voxel dimensions. This is a finite-volume interpretation in which the interface between two voxels carries the series resistance of the two half-voxels. The resulting system is
\begin{equation}
\mathbf{A}\boldsymbol{\Phi} = \mathbf{b},
\label{eq:system}
\end{equation}
where the off-diagonal entries of $\mathbf{A}$ are the pairwise conductances, each diagonal entry is the negative sum of the conductances incident on that node, and $\mathbf{b}$ encodes the injected current. The matrix $\mathbf{A}$ is therefore negative semidefinite. The system is solved in the equivalent form $-\mathbf{A}\boldsymbol{\Phi} = -\mathbf{b}$, in which the matrix is positive definite once the Dirichlet boundary is imposed.

Two boundary conditions are applied. The outer boundary is held at zero potential (Dirichlet), approximating a distant extracochlear return electrode. Voxels labeled as air are non-conducting. The zero-flux Neumann condition at tissue--air interfaces is imposed by omitting the corresponding edges during assembly, rather than by introducing auxiliary nodes. Both treatments preserve the symmetry of $\mathbf{A}$, which permits conjugate-gradient solution and guarantees the reciprocity property below.

Contact positions recovered from the CT are given to sub-voxel accuracy and rarely coincide with a node, so a contact is not represented as a single voxel. Injection and readout are instead distributed trilinearly over the eight nodes of the grid cell containing the contact, with weights $s_i$ summing to one. Corners falling on air or silicone nodes, which are constrained out of the system, are discarded and the remaining weights renormalized. A unit injection places $s_i$ at each retained corner, and the potential at a contact is read as $\sum_i s_i \Phi_i$ over its own corners. The same weights serve both operations, so the readout operator is the transpose of the injection operator, $\mathbf{Z}$ remains symmetric, and reciprocity is preserved exactly.

\begin{figure}[h]
\centering
\includegraphics[width=\textwidth]{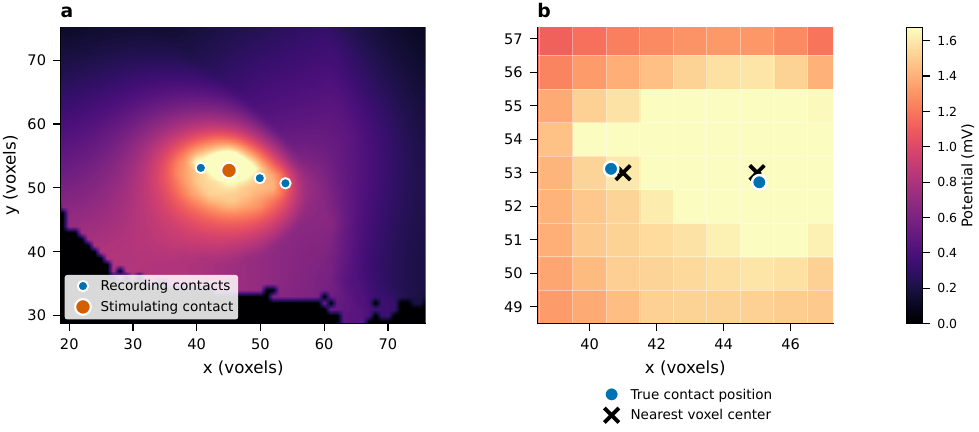}
\caption{(a) Simulated potential distribution on an axial slice through the
electrode array at the fitted parameters, on the reduced-resolution grid, for
stimulation at contact 11 (filled marker); open markers are the remaining
contacts in the slice.
(b) Detail at the scale of the reduced-resolution grid on which fitting is
performed. Injecting and reading at the nearest voxel center would displace
a contact by up to half a voxel, which the trilinear scheme of
Section~\ref{sec:forward-model} avoids}\label{fig:voltage}
\end{figure}

A SCINSEV measurement is simulated by injecting a unit current at one contact, solving for the potential, and reading the potential at the remaining contacts to form one column of the simulated transimpedance matrix $\mathbf{Z}$. Repeating across all stimulating contacts yields the full matrix. Throughout, a \emph{solve} denotes one solution of \eqref{eq:system} for a single stimulating contact, giving one column of $\mathbf{Z}$. An \emph{evaluation} denotes the set of solves required to score one candidate parameter vector. Fig.~\ref{fig:voltage} shows an example potential distribution and the sub-voxel contact placement used to read it out.

\subsection{Clinical SCINSEV measurements}
\label{sec:efi}

SCINSEV data were obtained from 43 ears of 43 patients implanted with 22-contact arrays (Cochlear Ltd.). All data were collected under IRB Approval Number 150417 and de-identified. 

Stimulation amplitude is set per patient and session rather than fixed across the cohort. Across these ears it ranged from 20.6 to 865.3~$\mu$A (median 125.3~$\mu$A). Recorded voltages are normalized by the delivered current, so the fitting target is a transimpedance rather than a raw voltage. Repeating across every stimulating contact yields a $22 \times 22$ matrix $\mathbf{Z}^{\mathrm{meas}}$, in which column $s$ holds the values recorded when contact $s$ is stimulated. This matrix is the fitting target.

\subsubsection{Reciprocity and the measurement noise floor}
\label{sec:noise-floor}

Under the model of Section~\ref{sec:forward-model} the simulated transimpedance matrix is symmetric, $Z_{rs}=Z_{sr}$. This reciprocity follows from linear, passive, steady-state conduction. It need not hold for the physical cochlea, where capacitive effects and electrode--tissue interface dynamics may introduce small asymmetries. We therefore read departures from reciprocity in the clinical SCINSEV as a measure of the mismatch between the idealized model and the measurement.

The reciprocity residual is
\begin{equation}
\varepsilon_{\mathrm{rec}} =
\frac{1}{N_{\mathrm{pairs}}}
\sum_{r<s}
\left| Z^{\mathrm{meas}}_{rs} - Z^{\mathrm{meas}}_{sr} \right|,
\label{eq:recip}
\end{equation}
computed over all contact pairs for which both entries are available. Across the cohort the median residual was 0.35\% in relative terms; in absolute terms the cohort mean was $\sigma_n = 0.00445$~mV. Throughout, $\sigma_n$ serves as a reference scale: discrepancies below it are indistinguishable from the reciprocity residual present in the measurements.

\subsubsection{Contact exclusion}
\label{sec:bad-contacts}

Contacts may yield unreliable measurements because of open or short circuits, poor contact with surrounding fluid, extracochlear placement, or localization error. Such contacts are identified once per patient during model preparation and the exclusion list is then fixed.

Three criteria are applied. First, contacts whose reciprocity residual exceeds three times the patient's median are flagged, indicating deviation from the symmetry expected under the forward model. Second, contacts whose measured profile is constant to numerical precision are flagged. These arise from sentinel placeholder values in the clinical export and would otherwise be read as valid. Third, contacts whose simulated response differs substantially between the full- and reduced-resolution models (Section~\ref{sec:acceleration}) are flagged. For these, the reduced-resolution approximation is unreliable, typically because downsampling alters the local conductive environment around the insulating silicone carrier.

A contact meeting any criterion is excluded. Exclusion is symmetric: the corresponding row and column are removed from both matrices before the objective is evaluated.

\subsection{Optimization pipeline}
\label{sec:pipeline}

\subsubsection{Objective function}
\label{sec:objective}

Let $\boldsymbol{\theta}$ denote the fitted compartment resistivities and $\mathbf{Z}(\boldsymbol{\theta})$ the simulated SCINSEV. Fitting minimizes
\begin{equation}
\mathcal{L}(\boldsymbol{\theta}) =
\sqrt{ \frac{1}{|\Omega|} \sum_{(r,s) \in \Omega}
\left( Z_{rs}(\boldsymbol{\theta}) - Z^{\mathrm{meas}}_{rs} \right)^2 } ,
\label{eq:objective}
\end{equation}
where $\Omega$ is the set of entries retained after excluding flagged contacts and diagonal entries ($r=s$), which are dominated by stimulation artifacts.

\subsubsection{Parameter identifiability and reduction}
\label{sec:param-reduction}

The model contains four resistivities, which are not equally recoverable from SCINSEV. To quantify identifiability, we computed the Jacobian $\mathbf{J}=\partial\mathbf{Z}/\partial\boldsymbol{\theta}$ by finite differences at three reference operating points per ear, spanning the admissible parameter range. For each parameter, we then measured the fraction of its sensitivity lying outside the span of the remaining columns,
\begin{equation}
\mathrm{rf}_k =
\frac{\left\| \mathbf{J}_k - P_{-k}\mathbf{J}_k \right\|}{\left\| \mathbf{J}_k \right\|},
\label{eq:rf}
\end{equation}
with $P_{-k}$ the least-squares projection onto the span of the remaining columns. Small $\mathrm{rf}$ indicates poor identifiability: the parameter's effect is approximable by a combination of the others.

Bone and soft tissue are nearly collinear in the Jacobian. Their sensitivity columns subtend a median $|\cos\angle| = 0.996$ (minimum 0.969) across all 43 ears and all three operating points. Bone therefore has little effect on the simulated SCINSEV that soft tissue cannot reproduce. Pooled across ears and operating points, the median residual fractions were $\mathrm{rf}_{\mathrm{stsv}} = 0.556$, $\mathrm{rf}_{\mathrm{mod}} = 0.214$, $\mathrm{rf}_{\mathrm{soft}} = 0.076$, and $\mathrm{rf}_{\mathrm{bone}} = 0.067$. ST/SV resistivity is the only strongly identifiable parameter, and bone the least identifiable of the four.

The modiolar residual fraction fell between these ($0.090$--$0.410$, median $0.214$) and varied with the operating point rather than across ears. Uniformly scaling all four resistivities left $\mathrm{rf}_{\mathrm{mod}}$ unchanged, whereas varying only the modiolus-to-soft ratio moved it. The variation therefore arises from relative current distribution rather than parameter magnitude, and no fitted value makes the modiolus reliably identifiable.

Accordingly, only ST/SV and soft-tissue resistivities were optimized. Bone was linked to soft through the fixed ratio $\rho_{\mathrm{bone}} = 20\rho_{\mathrm{soft}}$, and the modiolus fixed at the literature value 300~$\Omega\cdot$cm~\cite{eam-prior}. This retains the properties consistently recoverable from SCINSEV while avoiding optimization over parameters that are either largely redundant or strongly operating-point dependent. The reduction is evaluated in Section~\ref{sec:results}.

\subsubsection{Bayesian optimization}
\label{sec:bo}

Each evaluation of \eqref{eq:objective} requires a set of forward solves, so the number of evaluations dominates fitting cost. We use Bayesian optimization~\cite{bayesopt}, which builds a Gaussian process surrogate and selects each evaluation by maximizing an acquisition function over it. Relative to exhaustive search, this concentrates evaluations near the optimum. Unlike local gradient methods, it requires no derivatives with respect to the resistivities.

The search is initialized with 15 Latin hypercube points, followed by 30 sequential iterations using expected improvement~\cite{bayesopt}, for 45 evaluations per fit. The surrogate is a Gaussian process with a constant-scaled Mat\'ern kernel
($\nu = 5/2$)~\cite{gp-book} and an independent length scale per resistivity.
It is refitted by marginal-likelihood maximization after every evaluation, on
the internally standardized objective. Expected improvement is evaluated
with an exploration offset of $\xi = 0.01$ and maximized by L-BFGS-B from
random starts within the search bounds. The best evaluated point is
returned.

\subsubsection{Reducing the cost of each evaluation}
\label{sec:acceleration}

\paragraph{Reduced-resolution forward model.}
The forward model is evaluated on a grid downsampled by 0.402 in $x$ and $y$ and 0.404 in $z$ ($251 \times 251 \times 151 \rightarrow 101 \times 101 \times 61$), reducing nodes from $9.51\times10^{6}$ to $6.22\times10^{5}$, or 93.5\%. Because the field varies most rapidly near the contacts, the dominant downsampling error is displacement of each contact relative to the coarser voxel centers. This is addressed by the sub-voxel injection and readout of Section~\ref{sec:forward-model}, recovering the true contact position without refining the grid.

\paragraph{Symmetric assembly and conjugate gradient.}
Both boundary conditions preserve the symmetry of $\mathbf{A}$, so the system may be solved by conjugate gradient. The system is preconditioned with a degree-10 polynomial (Neumann-series) preconditioner~\cite{poly-precond}.

\paragraph{Solver tolerance.}
Solver tolerance was set to $10^{-5}$ on the residual norm. Against a high-precision reference ($\mathrm{tol}=10^{-8}$) across all 43 ears, the change in simulated SCINSEV at each recording contact was more than two orders of magnitude below $\sigma_n$ (median ratio $2.0\times10^{-3}$). The margin narrows quickly at looser tolerances. At $10^{-3}$ the discrepancy is only about 4.5-fold below $\sigma_n$, and at $10^{-2}$ it exceeds $\sigma_n$, so $10^{-5}$ was adopted.

\subsection{Experimental evaluation}
\label{sec:evaluation}

The pipeline introduces several approximations relative to a direct four-parameter fit at full resolution: reduced resolution, a halved set of stimulating contacts, a two- rather than four-dimensional parameter space, and surrogate-based rather than exhaustive optimization. Each is evaluated separately, then in combination, against the reference scales below.

\subsubsection{Reference scales}
\label{sec:reference-scales}

Two scales are used. The first is the measurement noise floor $\sigma_n$ (Section~\ref{sec:noise-floor}), derived from the per-patient reciprocity errors in our cohort. Any difference smaller than $\sigma_n$ is functionally inconsequential within our framework.

The second is the run-to-run variability of the fitting procedure, $\sigma_{\mathrm{seed}}$, obtained by repeating the fit on all 43 ears with 100 random initializations each. Because the objective is flat along partially degenerate directions, the fitted optimum is a region rather than a point, and $\sigma_{\mathrm{seed}}$ characterizes its extent. A difference below $\sigma_{\mathrm{seed}}$ is indistinguishable from the reproducibility of the pipeline, a weaker statement than falling below $\sigma_n$.

Across the cohort $\sigma_n$ was 0.00445~mV (mean) and $\sigma_{\mathrm{seed}}$ 0.0111~mV (mean, maximum 0.0183~mV).

\subsubsection{Evaluation of individual approximations}

\paragraph{Reduced-resolution forward model.}
For 43 ears, parameters fitted at reduced resolution were evaluated on the full-resolution model and compared against fitting directly at full resolution. Both arms use all 22 stimulating contacts, so the comparison isolates resolution. The comparison therefore measures whether the reduced model recovers an equivalent operating point, not whether it reproduces the same field. Fitted parameters may differ between resolutions either because the objectives have genuinely different minimizers, or because neither fit converges exactly. We also evaluated the landscapes of the objective function for both resolutions using $16 \times 16$ grids of the parameter spaces.

\paragraph{Electrode subset.}
Because the simulated matrix is symmetric, solving 11 of 22 stimulating contacts yields their columns directly and, by symmetry, their rows. Half the solves therefore recover 352 of the 462 off-diagonal entries. Fits using 11 and 6 stimulating contacts were compared against all 22 at fixed reduced resolution, isolating electrode count.

\paragraph{Parameter space reduction.}
Two-parameter fits were compared against fits with the modiolus additionally free, bone remaining linked to soft, across all 43 ears. The comparison was then restricted to the ten ears
with the highest $\mathrm{rf}_{\mathrm{mod}}$ at their own fitted operating
point ($>0.25$). These are the ears in which the modiolus is most nearly
identifiable, and therefore the least favorable cases for the reduction.

\subsubsection{Comparison of optimization strategies}

The Bayesian optimization procedure was compared against multi-resolution
grid search, gradient descent, Nelder--Mead, random search, and a sequential
coordinate strategy at matched evaluation budgets across all 43 ears. Grid
search used a $3\times3$ grid over the full bounds followed by four
successive grids centered on the current best with the step size halved at
each level. The sequential strategy swept each parameter over 11 values in
turn, with the other held at its current best. A second pass then covered
a $\pm25\%$ window around each. Both use 45 evaluations, matching the
Bayesian optimization budget. All strategy comparisons used the reduced-resolution model, isolating the optimizer.

\subsubsection{Comparison against exhaustive grid search}

The pipeline was additionally compared against exhaustive grid search ($16 \times 16$ over $(\rho_{\mathrm{stsv}}, \rho_{\mathrm{soft}})$, other parameters fixed and linked identically) at full resolution, on the 10 ears for which a full-resolution grid was available. This is distinct from the matched-budget comparison above, using 256 evaluations at full resolution rather than 45 at reduced.

\subsubsection{Implementation}

The forward model and pipeline were implemented in Python, with sparse linear algebra and the conjugate gradient solver on GPU via CuPy~\cite{cupy} (NVIDIA RTX A6000). Gaussian process regression used scikit-learn~\cite{sklearn}, with expected improvement implemented on top of the fitted surrogate. Reported fitting times exclude one-time preprocessing: tissue map construction, contact screening, and reduced-resolution model generation. This preprocessing is dominated by a single full-resolution evaluation over all 22 stimulating contacts ($\approx$2.1 min) and is amortized across all subsequent fits of that ear.

\section{Results}
\label{sec:results}

\subsection{Effect of individual approximations}

\subsubsection{Reduced-resolution forward model}
\label{sec:res-resolution}

Across all 43 ears, with both arms fitted over all 22 stimulating contacts,
the paired per-ear difference evaluated on full resolution was $-0.0002$~mV
(mean; median $-0.0005$~mV) with a standard deviation of 0.0120~mV
(Table~\ref{tab:resolution}). The two are equivalent within a margin of
$\sigma_{\mathrm{seed}}$ by two one-sided tests ($p<0.001$; 90\% CI
$[-0.0034, +0.0029]$~mV), and remain equivalent against the stricter margin
$\sigma_n$ ($p=0.014$). Coarsening the grid therefore costs nothing the
measurement can resolve.

\begin{table}[h]
\caption{Reduced- versus full-resolution fitting, all 43 ears, both arms over
all 22 stimulating contacts so that resolution is the only difference. Both
fitted parameter sets are scored on the full-resolution
objective}\label{tab:resolution}
\begin{tabular}{@{}lccc@{}}
\toprule
Parameters from & $n$ & Median RMSE (mV) & Mean $\pm$ SD (mV) \\
\midrule
Reduced-resolution fit & 43 & 0.0648 & $0.0790 \pm 0.0636$ \\
Full-resolution fit    & 43 & 0.0606 & $0.0792 \pm 0.0636$ \\
\midrule
Paired difference      & 43 & $-0.0005$ & $-0.0002 \pm 0.0120$ \\
\botrule
\end{tabular}
\footnotetext{Both scales for reference: $\sigma_n = 0.00445$~mV,
$\sigma_{\mathrm{seed}} = 0.0111$~mV.}
\end{table}

This near-equivalence was examined directly by mapping the
$(\rho_{\mathrm{stsv}}, \rho_{\mathrm{soft}})$ landscape with a
$16\times16$ exhaustive grid at both resolutions. Because a full-resolution
grid costs on the order of hours per ear, this was carried out on the ten
ears of the exhaustive-grid comparison (Section~\ref{sec:res-exhaustive}).
These span the cohort from the best-fitting to the worst-fitting ear. The
error surfaces correlated at $r=1.000$ (median across ears; minimum 0.998).
The minimizing grid point matched exactly in 8 of the 10 ears and differed
by one grid step in the other two. The two resolutions
therefore present the same objective at different discretizations, rather
than two objectives with different optima.

\subsubsection{Electrode subset}

Reducing the number of stimulating contacts left the fitted error
essentially unchanged. Fitting with 11 contacts gave a median RMSE of
0.0696~mV against 0.0691~mV for the full 22, a paired difference of
$+0.0004$~mV (mean) with a standard deviation of 0.0047~mV. Reducing further
to 6 contacts gave 0.0707~mV, again with a paired mean difference near zero
($+0.0005$~mV), but with the standard deviation roughly doubled to
0.0090~mV (Table~\ref{tab:electrodes}). Halving the stimulating contacts
therefore costs nothing the measurement can resolve, while quartering them leaves
the typical fit unchanged but makes individual fits appreciably less
stable. The pipeline accordingly solves 11 stimulating contacts, and every
result reported for it below is at that setting.

\begin{table}[h]
\caption{Effect of reducing the number of stimulating contacts, all 43 ears,
reduced-resolution model. Differences are paired against the 22-contact
fit.}\label{tab:electrodes}
\begin{tabular}{@{}lcccc@{}}
\toprule
Stimulating contacts & $n$ & Median RMSE (mV) & Mean RMSE (mV) & Paired diff.\ (mV) \\
\midrule
22 (all)          & 43 & 0.0691 & 0.0823 & --- \\
11 (every other)  & 43 & 0.0696 & 0.0827 & $+0.0004 \pm 0.0047$ \\
6 (every fourth)  & 43 & 0.0707 & 0.0828 & $+0.0005 \pm 0.0090$ \\
\botrule
\end{tabular}
\end{table}
\subsubsection{Parameter space reduction}

Freeing the modiolus while retaining the bone--soft link changed fitting
error by $+0.0024$~mV across all 43 ears (median; mean $+0.0049 \pm
0.0142$~mV) and by $-0.0050$~mV on the ten ears with the highest
$\mathrm{rf}_{\mathrm{mod}}$ (Table~\ref{tab:params}). Both are well inside
$\sigma_{\mathrm{seed}}$, so at a matched budget the third parameter is not
worth its cost either across the cohort or where the modiolus is most nearly
identifiable.

\begin{table}[h]
\caption{Effect of freeing the modiolar resistivity, with bone linked to soft
throughout. Both arms use the reduced-resolution objective at a matched budget
of 45 evaluations, and are scored on it. Positive values favor the
two-parameter fit}\label{tab:params}
\begin{tabular}{@{}lccc@{}}
\toprule
Cohort & $n$ & Median $\Delta$RMSE (mV) & Mean $\pm$ SD (mV) \\
\midrule
All ears & 43 & $+0.0024$ & $+0.0049 \pm 0.0142$ \\
Highest $\mathrm{rf}_{\mathrm{mod}}$ ($>0.25$) & 10 & $-0.0050$ & $-0.0036 \pm 0.0076$ \\
\botrule
\end{tabular}
\end{table}

The bone--soft ratio $k$ is not independently identifiable. The two are nearly collinear in the Jacobian (Section~\ref{sec:param-reduction}), so any fixed $k$ defines essentially the same fitting direction, with soft absorbing the difference. It is worth being precise about what this does and does not say. The residual fractions of the two are comparable ($\mathrm{rf}_{\mathrm{bone}} = 0.067$, $\mathrm{rf}_{\mathrm{soft}} = 0.076$) and their ranges overlap almost entirely, so a low residual fraction is not by itself a reason to eliminate bone rather than soft. What the collinearity establishes is that the pair contributes essentially one identifiable direction, so retaining either member and linking the other suffices. We retain soft tissue, the compartment in direct contact with the array, whose fitted value therefore carries the clearer physical interpretation. We use $k=20$, chosen to keep bone within a physically reasonable range given the search bounds on soft.

\subsection{Optimization}
\label{sec:res-optimization}

Across all 43 ears, Bayesian optimization achieved the lowest median and mean fitting error of the six strategies at a matched budget of 45 evaluations (Table~\ref{tab:strategy}).

\begin{table}[h]
\caption{Fitting error by optimization strategy, all 43 ears, reduced-resolution model, matched budget of 45 objective evaluations.}\label{tab:strategy}
\begin{tabular}{@{}lcc@{}}
\toprule
Strategy & Median RMSE & Mean $\pm$ SD RMSE \\
\midrule
Bayesian optimization & 0.069 & $0.082 \pm 0.057$ \\
Grid search (multi-res)          & 0.084 & $0.117 \pm 0.079$ \\
Sequential            & 0.085 & $0.096 \pm 0.060$ \\
Random search         & 0.111 & $0.122 \pm 0.059$ \\
Gradient descent      & 0.159 & $0.175 \pm 0.073$ \\
Nelder--Mead          & 0.184 & $0.202 \pm 0.057$ \\
\botrule
\end{tabular}
\end{table}

Grid search and the sequential strategy were the only competitive
alternatives, both within 0.016~mV of Bayesian optimization on median error.
On mean error Bayesian optimization's advantage over each exceeds
$\sigma_{\mathrm{seed}}$, and it also had the lowest standard
deviation of the six strategies.

\begin{figure}[h]
\centering
\includegraphics[width=0.75\textwidth]{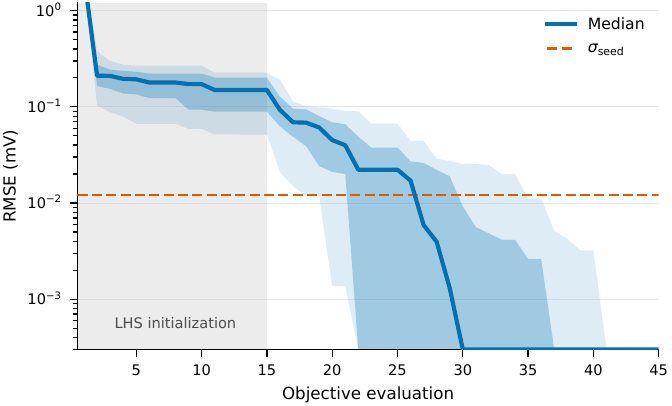}
\caption{Convergence of the fitting objective across all 43 ears, plotted
as the excess of the running-best objective over each ear's own final value.
The line is the cohort median; the dark and light bands are the interquartile
and 10th--90th percentile ranges. Shading marks the 15 Latin hypercube
initialization points. The running best falls within $\sigma_{\mathrm{seed}}$
of its final value after a median of 27 evaluations, supporting the budget of
45}\label{fig:convergence}
\end{figure}

Convergence curves (Fig.~\ref{fig:convergence}) show the running-best objective improved by less than $\sigma_{\mathrm{seed}}$ after a median of 27 evaluations, supporting the 45-evaluation budget.

\subsubsection{Comparison against exhaustive grid search}
\label{sec:res-exhaustive}

Against a full-resolution exhaustive grid ($16\times16$, 256 evaluations)
on the ten ears for which one was available, with both arms scored on the
full-resolution objective, Bayesian optimization reached a lower fitting
error on 5 of 10 ears. The paired median difference was $-0.0004$~mV
(mean $-0.0075$~mV, standard deviation 0.0180~mV). The pipeline is
non-inferior to the grid within a margin of $\sigma_{\mathrm{seed}}$
($p=0.006$, one-sided). The exhaustive grid's higher median error comes
from its 16-point spacing along each axis, not from its resolution: it can
only return a point on that lattice. This is why its median exceeds the
pipeline's while the paired difference is near zero.
The wall-clock comparison follows in Section~\ref{sec:res-overall}.

\subsection{Overall pipeline performance}
\label{sec:res-overall}

Combining all reductions, the pipeline fitted a patient-specific EAM in a
median of 2.6 minutes per ear. Two full-resolution references were used.
Against an independent Bayesian optimization at full resolution over all 22
stimulating contacts, across all 43 ears, the pipeline gave a paired
difference of $+0.0029$~mV (median; mean $+0.0025$~mV, standard deviation
0.0226~mV) at $37\times$ lower cost. Against an exhaustive $16\times16$ grid
search at full resolution, on the ten ears for which one was available, the
paired difference was $-0.0004$~mV (median; mean $-0.0075$~mV, standard
deviation 0.0180~mV) at $210\times$ lower cost (Table~\ref{tab:overall}).
Both differences are well inside $\sigma_{\mathrm{seed}}$, while both
standard deviations are roughly twice it, so the pipeline matches either
reference on average without reproducing it ear by ear.

\begin{table}[h]
\caption{Overall pipeline performance against two full-resolution
references. All fitting errors are scored on the full-resolution objective;
pipeline parameters are carried there by evaluating the fitted parameters on the full-resolution model. Times are per-ear wall clock on an NVIDIA RTX A6000.}
\label{tab:overall}
\begin{tabular}{@{}lcccc@{}}
\toprule
 & $n$ & Evaluations & Median RMSE (mV) & Time per ear \\
\midrule
\multicolumn{5}{@{}l}{\emph{Full cohort}} \\
Proposed pipeline (11 contacts) & 43 & 45 & 0.0625 & 2.6 min \\
Full-resolution BO, 22 contacts & 43 & 45 & 0.0606 & 1.6 h \\
Paired difference & 43 & --- & $+0.0029$ & $37\times$ faster \\
\midrule
\multicolumn{5}{@{}l}{\emph{Exhaustive-grid subset}} \\
Proposed pipeline (11 contacts) & 10 & 45 & 0.0453 & 2.6 min \\
Exhaustive grid ($16\times16$) & 10 & 256 & 0.0555 & 9.1 h \\
Paired difference & 10 & --- & $-0.0004$ & $210\times$ faster \\
\botrule
\end{tabular}
\end{table}
\section{Discussion}

Automation removes the operator dependence of manual parameter adjustment, making it feasible to fit an entire cohort under identical conditions and to re-fit when the model is revised.

The fitted resistivities should be read as effective bulk properties rather than tissue measurements, each absorbing any conduction pathway not explicitly represented. This reading is what licenses reducing the parameter space, since the objective is to reproduce measured electrical behavior rather than to recover physical conductivities.

The modiolus is the clearest case. SCINSEV constrains it weakly and
inconsistently (residual fraction 0.090 to 0.410 across ears and operating
points, median 0.214). When freed, it drifted to the low end of its range in
9 of 43 ears, and in 8 of those the ST/SV resistivity moved upward to
compensate (median $+22$~$\Omega\cdot$cm). The extra degree of freedom was
absorbing residual misfit rather than resolving modiolar tissue. Even on the
ten ears where it is most nearly identifiable, freeing it recovers only
0.005~mV, less than half of $\sigma_{\mathrm{seed}}$.

\begin{figure}[h]
\centering
\includegraphics[width=\textwidth]{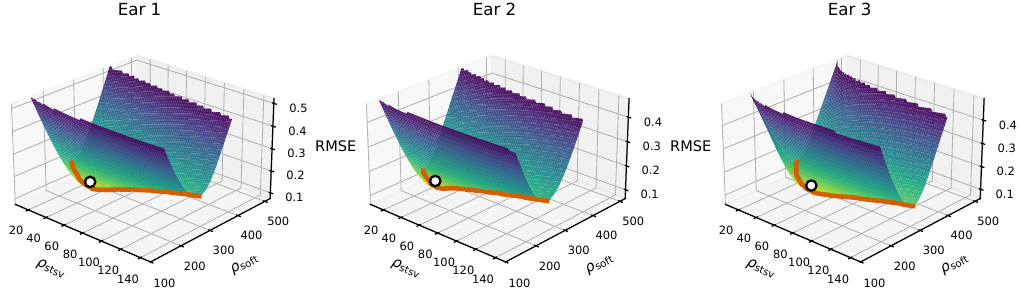}
\caption{Objective surface over the two free resistivities for three ears,
interpolated from the $16\times16$ reduced-resolution grid. The solid trace
marks the minimizing $\rho_{\mathrm{soft}}$ at each $\rho_{\mathrm{stsv}}$
and the open circle the global minimum. The
surface is truncated above the 60th percentile of fitting error to keep the
valley legible}\label{fig:manifold}
\end{figure}

The per-ear scatter throughout Section~\ref{sec:results} follows from the
shape of the objective. As the operating point moves, the direction of least
sensitivity rotates by a median of 44$^\circ$ between low- and
high-parameter reference points (maximum 72$^\circ$). The near-optimal set
is therefore a curved manifold rather than a straight valley
(Fig.~\ref{fig:manifold}). Two searches that agree in fitting error can thus
settle at different points along it, and we accordingly report agreement in
fitting error rather than in fitted parameter values.

Several limitations should be noted.
\label{sec:discussion-limitations}

\begin{itemize}
\item \textbf{Contact-potential-only validation.} All agreement reported
here is with respect to the potential at the contacts, since that is what
SCINSEV measures. Fields may differ further from the array, so downstream
applications such as neural activation modeling~\cite{anf-model} should be
evaluated on the full-resolution model.

\item \textbf{Tissue growth is not modeled.} Fibrous and osseous growth
around the array~\cite{fibrosis} has no term in the four-compartment model,
so affected ears are fitted without a representation of the structure
causing their residual. The poorest-fitting ear shows basal profile
distortion consistent with peri-electrode growth, suggesting its residual is
anatomical rather than a search failure. Extending the model to represent
growth is ongoing work, and detecting growth from a single recording is
addressed in companion work~\cite{tg-companion}.

\item \textbf{Single-device cohort.} All data were acquired with 22-contact
arrays (Cochlear Ltd.). The identifiability findings in particular may not
transfer to other array designs or stimulation configurations.
\end{itemize}

\section{Conclusion}

Fitting patient-specific EAMs to clinical SCINSEV has been limited by the
computational cost of the procedure and by its reliance on manual parameter
adjustment. We have shown that this cost can be substantially reduced by
coarsening the forward-model resolution, halving the number of stimulating contacts solved,
and reducing the number of fitted parameters. Each reduction was evaluated
against the reproducibility
of the fitting procedure, and their combined effect on accuracy remains
within it. The resulting automated pipeline matches an exhaustive
full-resolution grid search within that reproducibility,
in a median of 2.6 minutes per ear rather than 9.1 hours. This removes the
principal barrier to applying patient-specific EAMs at cohort scale, and
makes their use practical for retrospective study and for incorporation into
image-guided CI programming workflows.
\backmatter
\section*{Statements and Declarations}

\subsection*{Funding}
This research was supported by the National Institute on Deafness and Other
Communication Disorders grant number R01DC014037. This work is solely the
responsibility of the authors and does not represent the official views of
the NIH.

\subsection*{Competing Interests}
The authors have no relevant financial or non-financial interests to
disclose.

\subsection*{Ethics Approval}
Data collected in this study were authorized by the Vanderbilt IRB, IRB
Approval Number 150417, Vanderbilt University.

\subsection*{Consent to Participate}
Informed consent was obtained from all individual participants included in
the study.

\subsection*{Consent to Publish}
The manuscript contains no individually identifiable data.

\subsection*{Data Availability}
The clinical data analysed in this study are not publicly available, as they
derive from identifiable clinical imaging. Derived fitting results are
available from the corresponding author on reasonable request, subject to
institutional approval.

\subsection*{Code Availability}
The code is available from the corresponding author on reasonable request,
subject to institutional approval.

\subsection*{Author Contributions}
M.V.\ designed and implemented the optimization pipeline,
performed the experiments and analysis, and drafted the manuscript. E.B.\
contributed the preceding fitting pipeline and the forward-model
implementation it builds on. J.N.\ supervised the work and secured funding.
All authors reviewed and approved the final manuscript.


\begin{thebibliography}{99}

\bibitem{ci-outcomes}
Boisvert I, Reis M, Au A, Cowan R, Dowell RC (2020) Cochlear implantation
outcomes in adults: a scoping review. PLoS ONE 15(5):e0232421.
\url{https://doi.org/10.1371/journal.pone.0232421}

\bibitem{igcip}
Bratu EL, Sunderhaus L, Berg K, Dwyer R, Labadie RF, Gifford RH, Noble JH
(2021) Electrical stimulation overlap visualization for image-guided cochlear
implant programming. Proc SPIE 11598, Medical Imaging 2021: Image-Guided
Procedures, Robotic Interventions, and Modeling, 115981M.
\url{https://doi.org/10.1117/12.2581768}

\bibitem{efi-clinical}
Vanpoucke FJ, Zarowski AJ, Peeters SA (2004) Identification of the impedance
model of an implanted cochlear prosthesis from intracochlear potential
measurements. IEEE Trans Biomed Eng 51(12):2174--2183.
\url{https://doi.org/10.1109/TBME.2004.836518}

\bibitem{eam-prior}
Malherbe TK, Hanekom T, Hanekom JJ (2016) Constructing a three-dimensional
electrical model of a living cochlear implant user's cochlea. Int J Numer
Methods Biomed Eng 32(7):e02751. \url{https://doi.org/10.1002/cnm.2751}

\bibitem{anf-model}
Briaire JJ, Frijns JHM (2005) Unraveling the electrically evoked compound
action potential. Hear Res 205(1--2):143--156.
\url{https://doi.org/10.1016/j.heares.2005.03.020}

\bibitem{fdm-cochlea}
Agrawal V, Newbold C (2012) Computer modelling of the cochlea and the cochlear
implant: a review. Cochlear Implants Int 13(2):113--123.
\url{https://doi.org/10.1179/1754762811Y.0000000015}

\bibitem{bratu-lineage}
Bratu EL, Berg KA, DeFreese AJ, Gifford RH, Noble JH (2024) An adaptable model
for estimating patient-specific electrical properties of the implanted cochlea.
Proc SPIE 12928, Medical Imaging 2024: Image-Guided Procedures, Robotic
Interventions, and Modeling, 129280Q.
\url{https://doi.org/10.1117/12.3008731}

\bibitem{atlas-seg}
Noble JH, Labadie RF, Majdani O, Dawant BM (2011) Automatic segmentation of
intracochlear anatomy in conventional CT. IEEE Trans Biomed Eng
58(9):2625--2632. \url{https://doi.org/10.1109/TBME.2011.2160262}

\bibitem{elec-localization}
Zhao Y, Chakravorti S, Labadie RF, Dawant BM, Noble JH (2019) Automatic
graph-based method for localization of cochlear implant electrode arrays in
clinical CT with sub-voxel accuracy. Med Image Anal 52:1--12.
\url{https://doi.org/10.1016/j.media.2018.11.005}

\bibitem{prior-eam-solver}
Rybarczyk JA, Bratu EL, Labadie RF, Noble JH (2026) Simulation of apically
grounded cochlear implant stimuli using neural stimulation models. Int J
Comput Assist Radiol Surg.
\url{https://doi.org/10.1007/s11548-026-03702-3}

\bibitem{quasistatic}
Plonsey R, Heppner DB (1967) Considerations of quasi-stationarity in
electrophysiological systems. Bull Math Biophys 29(4):657--664.
\url{https://doi.org/10.1007/BF02476917}

\bibitem{bayesopt}
Jones DR, Schonlau M, Welch WJ (1998) Efficient global optimization of
expensive black-box functions. J Glob Optim 13(4):455--492.
\url{https://doi.org/10.1023/A:1008306431147}

\bibitem{gp-book}
Rasmussen CE, Williams CKI (2006) Gaussian Processes for Machine Learning.
MIT Press, Cambridge MA. \url{https://doi.org/10.7551/mitpress/3206.001.0001}

\bibitem{poly-precond}
Saad Y (2003) Iterative Methods for Sparse Linear Systems, 2nd edn. SIAM,
Philadelphia. \url{https://doi.org/10.1137/1.9780898718003}

\bibitem{cupy}
Okuta R, Unno Y, Nishino D, Hido S, Loomis C (2017) CuPy: a NumPy-compatible
library for NVIDIA GPU calculations. In: Proceedings of the Workshop on
Machine Learning Systems (LearningSys), 31st Conference on Neural Information
Processing Systems (NIPS).
\url{http://learningsys.org/nips17/assets/papers/paper_16.pdf}

\bibitem{sklearn}
Pedregosa F, Varoquaux G, Gramfort A, Michel V, Thirion B, Grisel O, Blondel M,
Prettenhofer P, Weiss R, Dubourg V, Vanderplas J, Passos A, Cournapeau D,
Brucher M, Perrot M, Duchesnay E (2011) Scikit-learn: machine learning in
Python. J Mach Learn Res 12:2825--2830.

\bibitem{fibrosis}
Somdas MA, Li PMMC, Whiten DM, Eddington DK, Nadol JB Jr (2007) Quantitative
evaluation of new bone and fibrous tissue in the cochlea following cochlear
implantation in the human. Audiol Neurotol 12(5):277--284.
\url{https://doi.org/10.1159/000103208}

\bibitem{tg-companion}
Vu M, Bratu E, Rybarczyk J, Noble J (2026) Single-visit screening for
intracochlear tissue growth validated against multi-visit electrical
measurement changes. Manuscript submitted for publication.
\end{thebibliography}
\end{document}